\documentclass[twocolumn,secnumarabic,showpacs,amssymb, amsmath,tightenlines,aps,prb]{revtex4-1}
\usepackage{longtable}
\usepackage{bm}
\usepackage{epsfig}
\usepackage{graphicx}

\begin{document}

\title[High-field metamagnetism in the antiferromagnet CeRh$_2$Si$_2$]
{High-field metamagnetism in the antiferromagnet CeRh$_2$Si$_2$}
\author{W. Knafo$^{1}$, D. Aoki$^{2}$, D. Vignolles$^{1}$, B. Vignolle$^{1}$, Y. Klein$^{1}$, C. Jaudet$^{1}$, A. Villaume$^{2}$, C. Proust$^{1}$, and J. Flouquet$^{2}$}

\address{$^{1}$ Laboratoire National des Champs Magn\'{e}tiques Intenses, UPR 3228, CNRS-UJF-UPS-INSA, 143 Avenue de Rangueil,
31400 Toulouse, France.}
\address{$^{2}$ Institut Nanosciences et Cryog\'{e}nie, SPSMS, CEA-Grenoble, 17 rue des Martyrs, 38054 Grenoble, France.}

\begin{abstract}

A study of the antiferromagnet CeRh$_2$Si$_2$ by torque,
magnetostriction, and transport in pulsed magnetic fields up to 50
Tesla and by thermal expansion in static fields up to 13
Tesla is presented. The magnetic field-temperature phase diagram
of CeRh$_2$Si$_2$, where the
magnetic field is applied along the easy axis $\mathbf{c}$, is deduced from these measurements. The
second-order phase transition temperature $T_{N}$ and the first-order phase transition temperature $T_{1,2}$ (=~36~K and 26~K at zero-field, respectively) decrease with increasing field. The field-induced antiferromagnetic-to-paramagnetic borderline $H_c$, which equals 26~T  at 1.5~K, goes from first-order at low temperature to second-order at high temperature. The magnetic field-temperature phase diagram is found to be composed of (at least) three different antiferromagnetic phases. These are separated by the first-order lines $H_{1,2}$, corresponding to $T_{1,2}$ at $H=0$, and $H_{2,3}$, which equals 25.5~T at 1.5~K. A maximum of the $T^2$-coefficient $A$ of the resistivity is observed at the onset of the high-field polarized regime, which is interpreted as the signature of an enhanced effective mass at the field-induced quantum instability. The magnetic field dependence of the $A$ coefficient in CeRh$_2$Si$_2$ is compared with its pressure dependence, and also with the field dependence of $A$ in the prototypal heavy-fermion system CeRu$_2$Si$_2$.

\end{abstract}


\maketitle

\section{Introduction}

CeRh$_2$Si$_2$ is a heavy-fermion antiferromagnet, crystallizing
in the ThCr$_2$Si$_2$ tetragonal structure, which can be driven to
a magnetic instability either by applying pressure \cite{ohashi03}
or magnetic field \cite{settai97,abe97}. It exhibits a second-order antiferromagnetic transition at the N\'{e}el temperature $T_{N}=36$~K and a first-order phase transition at $T_{1,2}=26$ ~K \cite{graf98,kawarazaki00}.  For temperatures $T_{1,2}\leq T\leq T_{N}$, the moments on the Ce sites order antiferromagnetically with wavevector (1/2,1/2,0). Below $T_{1,2}$ the antiferromagnetic structure is modified, the intensity of the (1/2,1/2,0) Bragg peak being strongly reduced while an additional (1/2,1/2,1/2) Bragg peak suddenly develops \cite{kawarazaki00}. De Haas - van Alphen experiments on this system at ambient pressure were interpreted in terms of localized $f$-electrons \cite{araki01}. Application of hydrostatic pressure induces a quantum phase transition to a paramagnetic Fermi liquid regime at a critical pressure of around 11 kbar \cite{ohashi03} and unconventional superconductivity emerges in the vicinity of the quantum phase transition below a critical temperature going up to $T^{max}_{SC}\approx0.4$~K \cite{movshovich96,araki02}. Above 11 kbar, an itinerant description of the $f$-electrons was proposed from studies of the Fermi surface \cite{araki01}. In Ref. \onlinecite{settai97} and \onlinecite{abe97}, the application of a magnetic field along the easy-axis $\mathbf{c}$ was found to induce two successive first-order transitions, around $H_c\simeq26$~T, between the low-field antiferromagnetic phase and a high-field polarized paramagnetic regime. At the metamagnetic transition, the magnetization jumps in two successive steps from 0.2 to 1.6~$\mu_B$/Ce (Ref. \onlinecite{settai97} and \onlinecite{abe97}).

In this article, we present a study of the properties of
CeRh$_2$Si$_2$ (at ambient pressure) in high magnetic fields
applied along the easy axis $\mathbf{c}$. Torque,
magnetostriction, and transport measurements have been carried out in pulsed magnetic fields up to 50~T, and thermal expansion measurements have been performed in static fields up to 13~T. This study enabled us to characterize precisely the magnetic field-temperature phase diagram of the system. We found that the transition temperatures $T_{N}$ and $T_{1,2}$ decrease with increasing magnetic field. The antiferromagnetic-to-paramagnetic polarization at the magnetic field $H_c$, which is a second-order transition at high temperature, becomes a first-order transition at low temperature where $\mu_0H_c$ equals 26~T at 1.5~K. Below 20~K, an additional first-order anomaly develops at a magnetic field $\mu_0H_{1,2}$, which equals 25.5~T at 1.5~K. These transition lines imply that the magnetic field-temperature phase diagram of CeRh$_2$Si$_2$ is composed of (at least) three antiferromagnetic phases. Fits of the low temperature resistivity show a strong and sharp enhancement of the quadratic coefficient $A(H)$ at the transition to the polarized regime. As well as antiferromagnetic fluctuations probably govern the pressure-induced criticality, ferromagnetic fluctuations might play a role at the field-induced instability. We compare the magnetic-field induced instability and previous studies of the pressure-induced instability in CeRh$_2$Si$_2$. Assuming that $A$ is proportional to the square of the average effective mass, which is dressed by the magnetic fluctuations, the magnetic field- and  pressure-driven enhancements of the mass are discussed. Finally, the properties of CeRh$_2$Si$_2$ are compared with those of the canonical example CeRu$_2$Si$_2$ of heavy-fermion metamagnetism.

Experimental details are given in Section \ref{exp}. The $(H,T)$ magnetic phase diagram inferred from our resistivity, torque, and thermal expansion measurements is presented in Section
\ref{phdiag}. Thermal expansion, magnetostriction, torque, and resistivity data are shown and analyzed in Sections \ref{thexp}, \ref{torque}, and \ref{transport}. In Section \ref{discussion}, we concentrate on the magnetic field-dependence of the quadratic resistivity term $A$, which is compared with its pressure-dependence and with the magnetic-field dependence of $A$ in CeRu$_2$Si$_2$.

\section{Experimental details} \label{exp}

Single-crystalline CeRh$_2$Si$_2$ samples were grown by the
Czochralski technique in a tetra-arc furnace. Their residual resistivity ratios of $\approx$~60 give evidence for the high quality of the crystals. Torque, magnetostriction, and transport experiments were performed up to 50~T at the pulsed magnetic field facility at the LNCMI-Toulouse. Thermal expansion measurements were made in
static magnetic fields up to 13~T. Torque measurements were performed using a commercial piezoresistive micro-cantilever developed by Seiko Instruments Incorporated. The sample was glued with Apiezon N grease to the cantilever. A one-axis rotating
sample holder allowed a small angle $\theta$ to be varied between the $\mathbf{c}$ direction and the magnetic field at ambient temperature. The variation of the piezoresistance of the cantilever was measured with a Wheatstone bridge with an AC excitation at a frequency of 70~kHz. Magnetostriction and thermal expansion were measured along the $\mathbf{c}$-axis using commercial strain gages from the company Kyowa$^\circledR$. A Wheatstone bridge allowed us to measure the difference between the variation of the length of the sample and a reference one (silicon). Magnetostriction and thermal expansion were measured at frequencies of 60 kHz and 20~Hz, respectively. For the resistance measurement, a current excitation of 10~mA at 60~kHz was applied along the $\mathbf{a}$-axis. The voltage (and a reference signal) was digitized using a high-speed digitizer and post-analyzed to perform the phase comparison. While three samples from the same batch have been measured and give similar results, the data presented here correspond to the sample which has the best geometric factor. Tiny and non-reproducible variations of the out-of-phase signal between two magnetic field pulses led to additional offsets in the resistance versus field data. These offsets were corrected so that the zero-field resistance from each resistance versus field data at a particular temperature corresponds to the resistance versus temperature data measured at zero-magnetic field. For all measurements, the magnetic field $\mathbf{H}$ was applied along $\mathbf{c}$ (with a small additional angle for the torque). Torque and magnetostriction measurements were performed in a longer-pulse magnet (55~ms of rising field and 300~ms of falling field) than the resistivity measurements (26~ms of rising field and 110~ms of falling field). We show only data collected during the decreasing field part of the magnetic field pulse.

\section{Magnetic field-temperature phase diagram} \label{phdiag}

\begin{figure}[b]
    \centering
    \epsfig{file=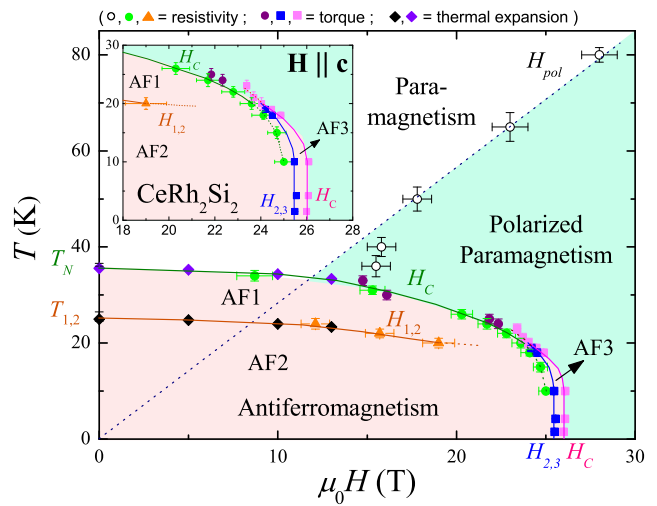,width=85mm}
    \caption{(Color online) Magnetic field-temperature phase diagram of CeRh$_2$Si$_2$, with $\mathbf{H}\parallel\mathbf{c}$, obtained from resistivity, torque, and thermal expansion. The insert focuses on the low-temperature part of the phase diagram.}
    \label{phdiag_all}
\end{figure}

Fig. \ref{phdiag_all} shows the magnetic field-temperature phase
diagram of CeRh$_2$Si$_2$, for $\mathbf{H}\parallel\mathbf{c}$,
constructed from our resistivity, torque, and thermal expansion measurements presented in Sections \ref{thexp}, \ref{torque}, and \ref{transport}. The antiferromagnetic phase, which develops at zero-field below $T_N=36$~K, is destabilized in magnetic fields $H$ higher than $H_c$, which equals 26~T at $T=1.5$~K. $H_c$ corresponds to a field-induced transition to a paramagnetic polarized regime with a strong polarization of the Ce moments \cite{settai97,abe97}. The antiferromagnetic-to-paramagnetic borderline $T_N$ (or equivalently $H_c$) goes from second-order above $\approx20$~K (and below 24~T) to first-order below $\approx20$~K (and above 24~T). Inside the antiferromagnetic phase two different magnetic transitions correspond to the first-order transitions at $T_{1,2}$ (or equivalently $H_{1,2}$) and $H_{2,3}$. They separate at least three antiferromagnetic phases, noted here AF1, AF2, and AF3. The transition temperature $T_{1,2}$, which equals 26~K at zero-magnetic field (or equivalently the magnetic field $H_{1,2}$), separates the antiferromagnetic phases AF1 and AF2. Both $T_{N}$ and $T_{1,2}$ decrease with increasing magnetic field. The transition line $H_{2,3}$ corresponds to a field-induced transition between the phases AF2 and AF3, this last phase being stable in a very narrow field-range of about 0.5~T. Torque measurements (see Section \ref{torque}) show that $H_{2,3}$ and $H_c$, which are distinct at low temperature, merge at about (24~T,~20~K). Our data are compatible with the presence of a critical point at around (24~T,~20~K) where all the antiferromagnetic transition lines would merge. However, the temperature uncertainty and the limited resolution of our experiments in pulsed fields (see  Sections \ref{thexp}, \ref{torque}, and \ref{transport}) do not allow us to conclude if this critical point really exists. Further measurements in static high magnetic fields would be necessary to check more carefully how the different transition lines behave in the proximity of the point (24~T,~20~K) and to test if they merge in a unique critical point. Finally, the high-temperature part of the phase diagram is characterized by a crossover at a magnetic field $H_{pol}$, defined here using resistivity data (see  Section \ref{transport}), between the low-field antiferromagnetically correlated phase and the high-field polarized phase. $H_{pol}$ increases linearly with $T$ or, equivalently, the characteristic temperature $T_{pol}$ of the high-field polarized state is proportional to $H$.

\section{Thermal expansion and magnetostriction} \label{thexp}

\begin{figure}[b]
    \centering
    \epsfig{file=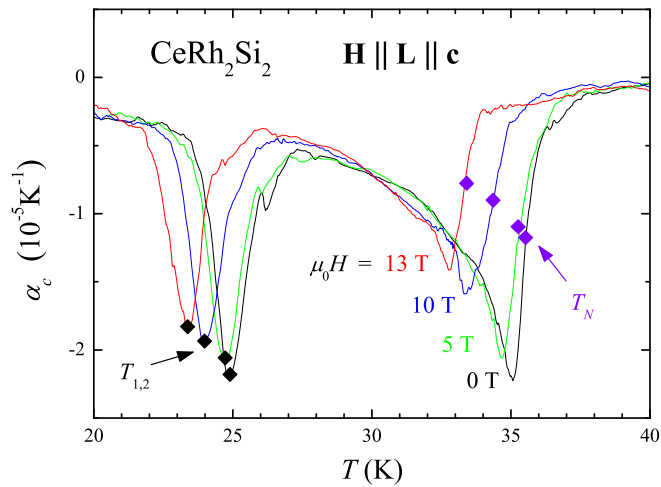,width=87mm}
    \caption{Thermal expansion coefficient versus temperature of CeRh$_2$Si$_2$ measured for magnetic fields $\mu_0H=0$~T, 5~T, 10~T, and 13~T applied along $\mathbf{c}$.}
    \label{alpha}
\end{figure}

The temperature dependence of the thermal expansion coefficient
$\alpha_c=1/L_c\times\partial L_c/\partial T$, where $L_c$ is the length of the sample along $c$, is plotted in Fig. \ref{alpha} for $20\leq T\leq40$~K and magnetic fields $\mathbf{H}\parallel \mathbf{c}$ of 0~T, 5~T, 10~T, and 13~T. Thermal expansion at zero-magnetic field indicates the presence of two phase transitions, a step-like anomaly in $\alpha_c(T)$ is found at the second-order phase transition temperature $T_{N}=35.5\pm0.1$~K (defined at the extremum of slope of $\alpha_c(T)$) and a symmetric negative peak is found at the first-order transition temperature $T_{1,2}=25\pm0.1$~K (defined at the minimum of $\alpha_c(T)$) \cite{notetransitions}. Our zero-field data are in good agreement with previous thermal expansion \cite{araki98,villaume07} and specific heat measurements \cite{boursier05}, which also indicated the second-order nature of $T_{N}$ and the first-order nature of $T_{1,2}$ (see also [\onlinecite{notetransitions}]). However, the relative change in length ($\Delta L_c/L_c\simeq1.7\times10^{-4}$) between 5 K and $T_{N}$ corresponding to the zero-field thermal expansion coefficient $\alpha_c$ plotted in Fig. \ref{alpha} is 30~$\%$ smaller than the variation of around $2.5\times10^{-4}$ reported using absolute capacitive dilatometry technique \cite{settai97,araki98}. The strain gauge is thus not perfectly coupled to the sample. The efficiency of the coupling is estimated to 70~$\%$. For this reason, the anomalies in $\alpha_c(T)$ reported here at  $T_N$ and $T_{1,2}$ are smaller than those from Ref. \onlinecite{araki98} and \onlinecite{villaume07}. As shown in Fig. \ref{alpha}, both $T_{N}$ and $T_{1,2}$ decrease when a magnetic field is applied along $\mathbf{c}$. Since the anomalies at $T_{N}$ and $T_{1,2}$ in $\alpha_c(T)$ are both negative, the Ehrenfest and Clapeyron relations respectively imply that, in magnetic fields $0\leq\mu_0H\leq13$~T parallel to $c$, uniaxial pressures applied along $c$ would decrease both $T_{N}$ and $T_{1,2}$.

\begin{figure}[t]
    \centering
    \epsfig{file=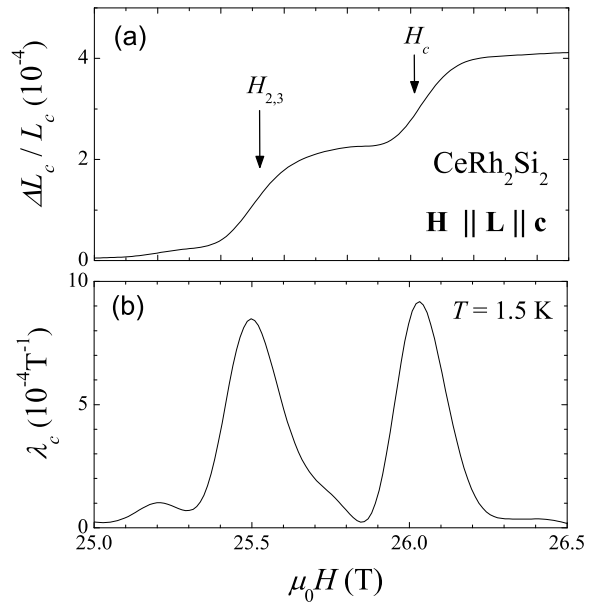,width=77mm}
    \caption{(Color online) Magnetic field-dependence at $T=1.5$~K and for $\mathbf{H}\parallel\mathbf{c}$, (a) of the relative length $\Delta L_c/L_c$ of CeRh$_2$Si$_2$ and (b) of the related magnetostriction coefficient $\lambda_c$.}
    \label{lambda}
\end{figure}

Fig. \ref{lambda} (a) shows a plot of the magnetic field-variation ($\mathbf{H}\parallel\mathbf{c}$) of $\Delta L_c/L_c$ measured at $T=1.5$~K. A two-step like increase of $\Delta L_c/L_c$ is induced at the first-order transitions $H_{2,3}\simeq25.5$~T and $H_c\simeq26$~T leading to two well-defined peaks in the magnetostriction coefficient $\lambda_c=1/L_c\times\partial L_c/\partial(\mu_0H)$ (Fig. \ref{lambda} (b), see also [\onlinecite{notetransitions}]). The two steps in length variation $(\Delta L_c/L_c)_1\simeq(\Delta L_c/L_c)_2\simeq2\times10^{-4}$ measured at $H_{2,3}$ and $H_c$ recall those, equal to $\Delta M_1\simeq\Delta M_2\simeq0.7$~$\mu_B$, observed in the magnetization at $H_{2,3}$ and $H_c$ (Ref. \onlinecite{settai97} and \onlinecite{abe97}). Taking into account the 30 $\%$ reduction in sensitivity of the length variation, we can estimate the real length variations at $H_{2,3}$ and $H_c$ by $(\Delta L_c/L_c)_{1,2}^{real}\simeq3\times10^{-4}$.

\section{Torque} \label{torque}

\begin{figure}[t]
    \centering
    \epsfig{file=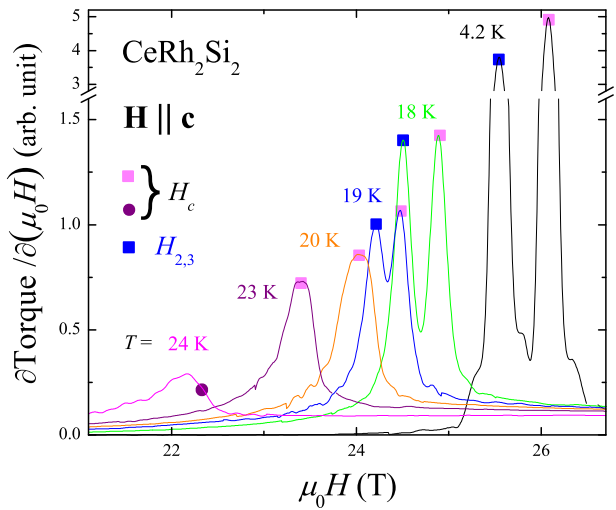,width=82mm}
    \caption{(Color online)  Magnetic field-derivative of the torque in CeRh$_2$Si$_2$ versus magnetic field for temperatures $T\leq24$~K and magnetic fields along $\mathbf{c}$.}
    \label{dtorquedH}
\end{figure}

Fig. \ref{dtorquedH} shows a plot of the field-derivative of the
torque versus magnetic field of CeRh$_2$Si$_2$ at temperatures
between 4.2~K and 24~K. The torque signal is proportional to
$MH$sin$\theta$ where $M$ is the magnetization and $\theta$ is a
small angle between the magnetic field $\mathbf{H}$ and the easy
axis $\mathbf{c}$ of the sample. The field-induced polarization of the system is accompanied at 4.2~K by two successive steps in the torque, which lead to two well-defined maxima in the field-derivative of the torque at the first-order transitions fields $H_{2,3}\approx25.5$~T and $H_c\approx26$~T. Our torque data are in good agreement with our magnetostriction data (see Section \ref{thexp}), and also with magnetization measurements performed by Settai et al. \cite{settai97} and Abe et al. \cite{abe97}, in which two first-order transitions were reported at similar magnetic fields. From Fig. \ref{dtorquedH}, it is clear that the two transitions $H_{2,3}$ and $H_c$ merge at about 20~K into a single first-order transition $H_c$. We note that $H_{2,3}$ and $H_c$ were found to be distinct up to 24~K in the magnetization data from Settai et al. \cite{settai97}. In our data, a first-order like anomaly at $H_c$ can be seen up to 23~K. This is characterized by a symmetric positive anomaly in the field-derivative of the torque (Fig. \ref{dtorquedH}). For $T\geq24$~K, an asymmetric step-like anomaly as opposed to the symmetric maximum observed at lower temperatures is observed at $H_c$ in the field-derivative of the torque. This anomaly indicates that the transition is of second-order \cite{notetransitions}.

\section{Resistivity} \label{transport}

\begin{figure}[b]
    \centering
    \epsfig{file=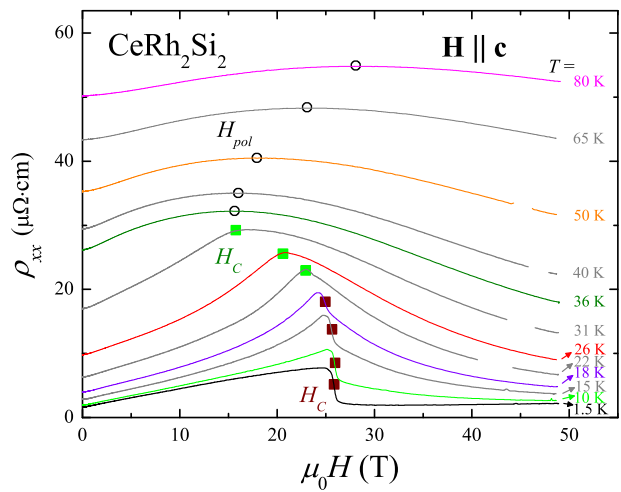,width=82mm}
    \caption{(Color online)  Resistivity $\rho_{xx}$ versus magnetic field $H$ in CeRh$_2$Si$_2$, for $\mathbf{H}\parallel\mathbf{c}$ and $1.5\leq T\leq 80$~K.}
    \label{RH}
\end{figure}

Fig. \ref{RH} shows measurements of the resistivity $\rho_{xx}$ versus $H$ of CeRh$_2$Si$_2$ for magnetic fields up to 50~T and temperatures between 1.5~K and 80~K. At low temperatures, a step-like anomaly is observed at the critical field $\mu_0H_c=25.9$~T ($H_c$ is defined at the extremum of slope of $\rho_{xx}(H)$). This anomaly corresponds to the antiferromagnetic-to-paramagnetic borderline of the system. Our resistivity data show only one first-order transition to the polarized regime, while torque and magnetostriction data permitted us to observe two successive transitions at $H_{2,3}$ and $H_c$ (Sections \ref{thexp} and \ref{torque}). The strong change of resistivity between the antiferromagnetic and the polarized phases is associated with a reconstruction of the Fermi surface as recently detected by quantum oscillations in measurements of the torque in high static magnetic fields \cite{demuer09}. The transition field $H_c$ decreases with increasing temperature and reaches zero at $T_{N}(H=0)=36$~K. Below 20~K, the anomaly at $H_c$ has an asymmetric step-like shape and can be considered as the signature of a first-order transition. At 20~K, the shape of $\rho_{xx}(H)$ is almost symmetric and a change of slope at a magnetic field of around 24 T has replaced the step-like anomaly observed at low temperature. For $20\leq T\leq36$~K, a second-order-like change of slope of $\rho_{xx}(H)$ can be defined at the antiferromagnetic transition field $H_c$. The $\rho_{xx}$ versus $H$ data plotted in Fig. \ref{RH} indicate that, at around (24~T,~20~K), the antiferromagnetic-to-paramagnetic transition in CeRh$_2$Si$_2$ goes from first-order at low temperature to second-order at high temperature. For $T>36$~K, a broad maximum of $\rho_{xx}(H)$ is found at a magnetic field $H_{pol}$, which increases with $T$. This anomaly is attributed to the crossover between the low-field paramagnetic
regime, where antiferromagnetic correlations dominate, and the high-field polarized paramagnetic regime. The initial positive slope of the magnetoresistivity is attributed to antiferromagnetic correlations. The persistence of a positive slope in $\rho_{x,x}(H)$ at 80~K implies that antiferromagnetic correlations develop above 80~K. If the Ce ions would be independent, i.e., only subject to single-site phenomena as in the Kondo effect, a negative slope of the magnetoresistivity would be expected at all magnetic fields.

\begin{figure}[b]
    \centering
    \epsfig{file=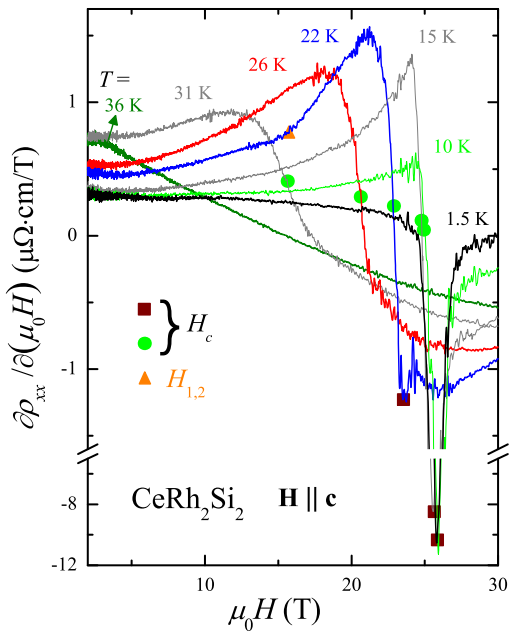,width=68mm}
    \caption{(Color online) Magnetic field-derivative of the resistivity $\partial\rho_{xx}/\partial(\mu_0H)$ versus $H$ in     CeRh$_2$Si$_2$ for $\mathbf{H}\parallel\mathbf{c}$ and $1.5\leq T\leq36$~K.}
    \label{dRdH}
\end{figure}

The derivative of the resistivity $\partial\rho_{xx}/\partial(\mu_0H)$ is plotted in Fig. \ref{dRdH} for $\mu_0H\leq30$~T and 1.5 $\leq T\leq$ 36~K. Below 10~K, a well-defined minimum of $\partial\rho_{xx}/\partial(\mu_0H)$ is obtained at the first-order transition field $H_c$. For $T>10$~K, the size of the first-order-like anomaly in $\partial\rho_{xx}/\partial(\mu_0H)$ decreases with increasing $T$, being progressively replaced by a step-like anomaly in $\partial\rho_{xx}/\partial(\mu_0H)$.
Between 26~K and 36~K, a clear step-like anomaly is observed in
$\partial\rho_{xx}/\partial(\mu_0H)$ at $H_c$, which is defined at the extremum of slope of $\partial\rho_{xx}/\partial(\mu_0H)$). This step coincides with the second-order nature of the transition. The fact that both kinds of anomalies, i.e., a step and a minimum, can be defined in all $\partial\rho_{xx}/\partial(\mu_0H)$ versus $H$ curves for $10\leq T\leq24$~K shows that the change between the low- and the high-temperature regimes is not so well-defined in our resistivity data. To understand these features, it would be interesting to investigate the transport properties of other systems where a magnetic transition also goes from second- to first-order when the temperature is lowered. The size of the step in $\partial\rho_{xx}/\partial(\mu_0H)$ reaches a maximum at $T=20$~K, which may be related to enhanced magnetic fluctuations at around (24~T, 20~K), their intensity decreasing below 20~K when the transition becomes first-order. In Fig. \ref{dRdH}, an additional anomaly can be observed in $\partial\rho_{xx}/\partial(\mu_0H)$ at the magnetic field $H_{1,2}$, for the temperatures $20\leq T\leq24$~K. This anomaly corresponds to the transition between the antiferromagnetic states AF1 and AF2, which occurs at $T_{1,2}=26$~K in zero-field. Since $T_{1,2}$ is a first-order transition, an extremum of $\partial\rho_{xx}/\partial(\mu_0H)$, similarly to the one observed at $H_c$, should be expected at $H_{1,2}$. A first-order-like anomaly at $H_{1,2}$ was obtained by Levy et al. using static high magnetic fields \cite{demuer09}, but this anomaly is probably hidden here by a broadening of the transition due to the pulsed nature of the magnetic field.

\begin{figure}[t]
    \centering
    \epsfig{file=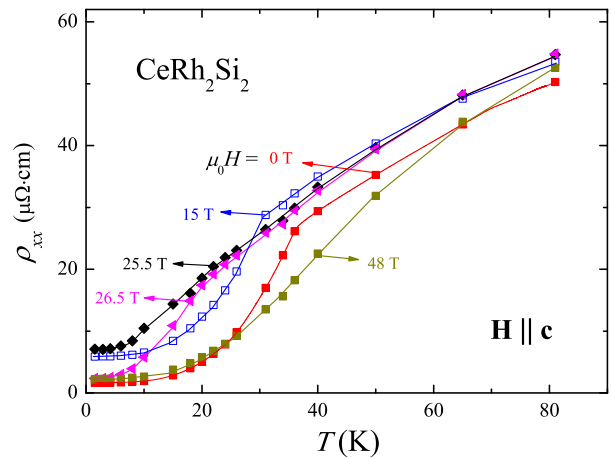,width=80mm}
    \caption{(Color online) Resistivity versus temperature $\rho_{xx}(T)$ of CeRh$_2$Si$_2$, for magnetic fields $\mathbf{H}\parallel\mathbf{c}$ of 0, 15, 25.5, 26.5 and 48~T.}
    \label{RT}
\end{figure}

\begin{figure}[t]
    \centering
    \epsfig{file=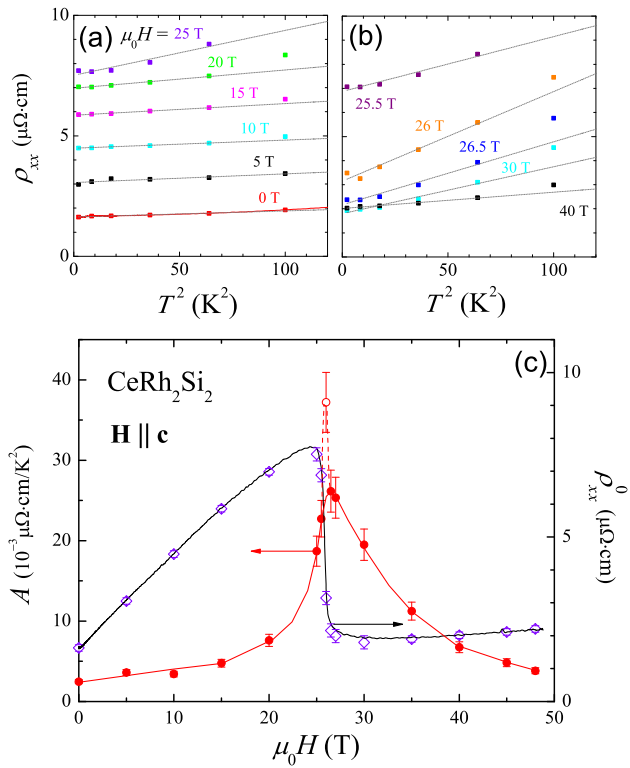,width=83mm}
    \caption{(Color online) Resistivity of CeRh$_2$Si$_2$ in a $\rho_{xx}$ shown versus $T^2$, for $\mathbf{H}\parallel\mathbf{c}$ and a) $H\leq25$~T, b) $H\geq25.5$~T. Dotted lines correspond to fits to the data, for $T\leq8$~K, by $\rho_{xx}=\rho^0_{x,x}+A T^2$. c) Plots of the quadratic coefficient $A$ and of the residual resistivity $\rho^0_{x,x}$ as a function of $H$. Red closed circles show the coefficient $A$ extracted for $H\neq H_c$ (where a quadratic fit sounds reasonable), while a red open circle is used for $A$ extracted at $H_c$ (where a quadratic fit is probably not justified). The errors bars come from the uncertainty in the numerical fits to the data. Red (full and dotted) lines are guides to the eyes for the field variation of $A$. The black line shows the field-dependence of $\rho_{xx}$ measured at 1.5 K.}
    \label{AR0H_RT2}
\end{figure}

Fig. \ref{RT} shows the temperature dependence of $\rho_{xx}$
at different magnetic fields. The curve at
zero-field was measured separately, while the data in magnetic
field have been extracted from the $\rho_{xx}$ versus $H$ data
plotted in Fig. \ref{RH}. No anomaly is seen at 25.5~T in contrast with the clear kinks at 36~K and 25~K in zero-field and at 15~T, respectively, which correspond to the transition line $T_N$ (cf. the phase diagram in Fig. \ref{phdiag_all}). However, the $T$-dependence of $\rho_{xx}$ under 25.5~T over the wide $T$-window from above 20~K down to low temperatures looks quite anomalous. At low temperature,
$\rho_{xx}$ drops suddenly through $H_c$ and the system becomes highly polarized paramagnetically. Plots of $\rho_{xx}$ versus $T^2$ are shown in Fig. \ref{AR0H_RT2} (a) for magnetic fields below 25~T and in Fig. \ref{AR0H_RT2} (b) for magnetic fields above 25.5~T.
In these plots, the dotted lines correspond to the fits to the
data by $\rho_{xx}=\rho^0_{xx}+A T^2$ made for $T\leq8$ K. In
Fig. \ref{AR0H_RT2} (c), the quadratic coefficient $A$ and the
low-temperature resistivity $\rho^0_{xx}$ extracted from the fits
are presented. The step-like anomaly in $\rho^0_{xx}$ of course governs the behavior of $\rho_{xx}$ at very low temperature (Fig.
\ref{RH}). A step-like anomaly was also reported in the $\rho_{xx}$
versus $p$ data at a critical pressure of around 10~kbar
corresponding to the quantum phase transition to a paramagnetic
regime \cite{ohashi03,araki02}. In the low-field antiferromagnetic state, quantum oscillations and band calculations based on the $4f$-localized model have shown that the Fermi surface is multiple connected \cite{araki01}. Thus, the orbital magnetoresistance is expected to be saturated and field-independent, at least below the critical field $H_c$. A Fermi surface reconstruction probably occurs at $H_c$, as indicated by the sudden variation of $\rho^0_{xx}$. The slight increase of $\rho^0_{xx}$ observed above $H_c$ (see Fig. \ref{AR0H_RT2} (c)) could result from an orbital effect due to a reconstruction of the Fermi surface in closed orbits at high fields. To support this hypothesis, the high-field condition $\omega_c\tau>1$, where $\omega_c$ is the cyclotron frequency and $\tau$ the lifetime of the electron, should be fulfilled. This could be the case here, since the value of $\rho^0_{xx}\simeq1.6~\mu\Omega\cdot cm$ at zero-field obtained on our sample is not so far from the value of $1.3~\mu\Omega\cdot cm$ measured on the sample studied in Ref. \onlinecite{araki01}, for which quantum oscillations were reported at 30~mK. On the other hand, the increase of the magnetoresistivity observed above 30 T does not follow a clear $\Delta\rho/\rho\sim(\omega_c\tau)^2\sim H^2$ behavior, so that the high-field condition $\omega_c\tau>1$ could also be not yet fulfilled. Positive magnetoresistivity could also result from disorder effects as described in Ref. \onlinecite{ohkawa90}. Further experiments on samples of different qualities are required to solve this problem. A maximum of $A$ is obtained at the transition $\mu_0H_c=26$~T to the polarized regime. The maximal value of $A$, which is a factor 15 bigger than the one found at zero magnetic field, is probably affected by the sudden step in $\rho^0_{xx}$ in the narrow region around $H_c$. A Fermi surface reconstruction presumably controls this step, around which the resistivity might not be dominated by the collisions of the quasiparticules. Thus, the close region around $H_c$ might not be described within the Kadowaki-Woods approach \cite{kadowaki86}. A Kadowaki-Woods approach is appropriate only when inelastic scattering processes, i.e., mechanisms related to the magnetic fluctuations of the $f$-electron moments in heavy-fermion systems, dominate the electronic effective mass and control the temperature dependence of the resistivity. Here the $A$ coefficient might be proportional to the square root of the average effective mass only in magnetic fields below 25.5~T or above 26~T, assuming that the carrier density does not vary significantly. In these field ranges, $A$ passes through a broad maximum at $H_c$, being 10 times bigger than its value at zero magnetic field. Resistivity data are thus consistent with an enhancement of the heavy-fermion effective mass by a factor of 3 at $H_c$. Because of this factor 3, fits close to $H_c$ should have been made in a temperature window 3 times smaller than the fit made at zero-field. However, our sensitivity in pulsed magnetic fields does not allow to perform such an analysis.

\section{Discussion} \label{discussion}

\begin{figure}[b]
    \centering
    \epsfig{file=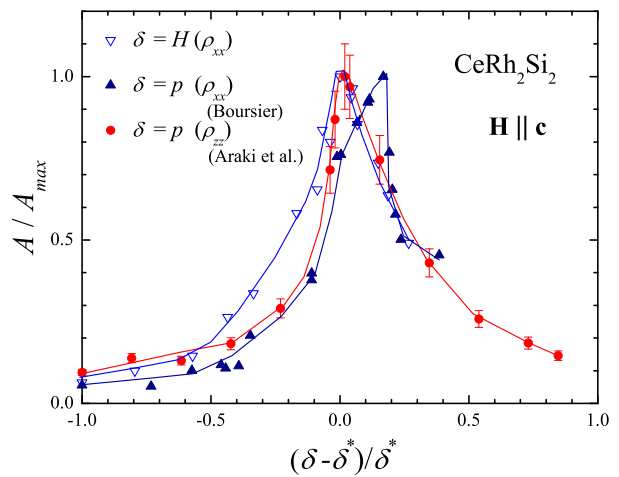,width=80mm}
    \caption{(Color online) Comparison of the magnetic field- and
    pressure-dependences \cite{araki02,boursier05} of the normalized quadratic coefficient $A/A_{max}$ of the low-temperature resistivity of CeRh$_2$Si$_2$ versus $(\delta-\delta^*)/\delta$.
    $\delta$ is the magnetic field or the pressure and
    $\delta^*$ its critical value.}
    \label{AHAp}
\end{figure}

\begin{figure}[b]
    \centering
    \epsfig{file=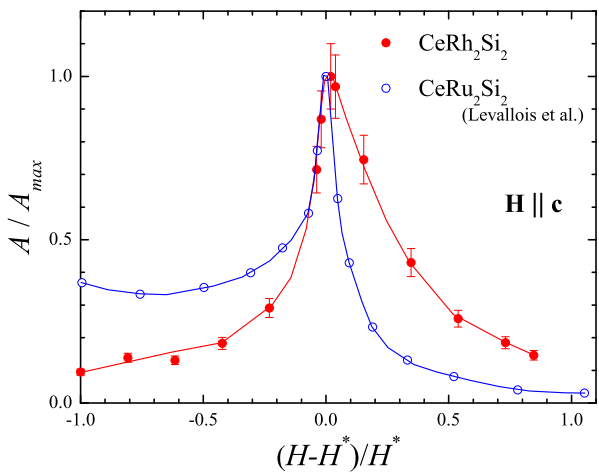,width=79mm}
    \caption{(Color online) Comparison of the magnetic field-dependences of the quadratic coefficient $A$ of the low-temperature resistivity, in a $A/A_{max}$ versus $(H-H^*)/H$ plot, for the heavy fermions CeRh$_2$Si$_2$ and CeRu$_2$Si$_2$ (Ref. \onlinecite{levallois09}).}
    \label{AHcomp}
\end{figure}

As shown in Fig. \ref{AR0H_RT2} (c), the $H$-enhancement of $A$ occurs over a broad magnetic field window, of about 10~T, while the two transitions at $H_{2,3}$ and $H_c$ are only separated by 0.5~T. Their experimental width is less than 0.1~T
through the first-order metamagnetic transitions. A further comparison of the pressure- \cite{araki02,boursier05} and magnetic field-variations of $A/A_{max}$, as a function of $(p-p_c)/p_c$ and $(H-H_c)/H_c$ (see Fig. \ref{AHAp}), leads to the remarkable result that their variations are almost comparable. Because the $A$ coefficient extracted at 26 T (see Fig. \ref{AR0H_RT2}) is surely affected by the sudden variation of $\rho^0_{xx}$ between 25.5 and 26 T it is not included in Fig. \ref{AHAp} (as well as in Fig. \ref{AHcomp}).

\begin{figure}[t]
    \centering
    \epsfig{file=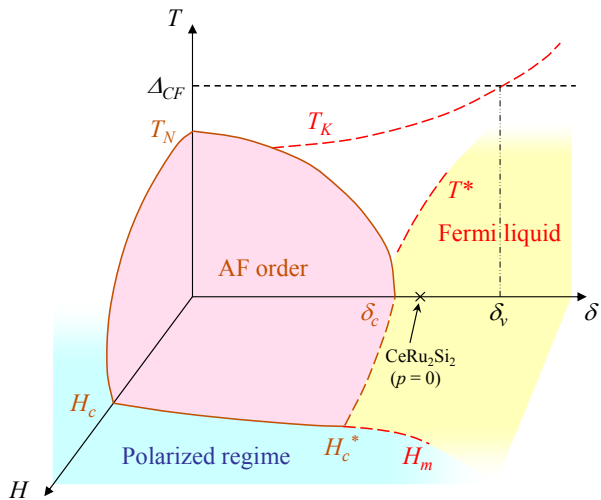,width=79mm}
    \caption{(Color online) Schematic magnetic field-pressure or doping-temperature phase diagram of the prototypal heavy-fermion system CeRu$_2$Si$_2$, when a magnetic field is applied along $\mathbf{c}$.}
    \label{CeRuSi_diag}
\end{figure}

A similarity between the critical values reached at $p_c$ and
$H_c$ has already been observed in the well-documented
heavy-fermion series CeRu$_2$Si$_2$, which is composed of
Ising-type magnetic centers on the cerium sites, and where quantum
criticality can be reached via lanthanum doping (fictitious
negative pressure) or via pressure tuning. The $(T,\delta,H)$
phase diagram of CeRu$_2$Si$_2$ is represented schematically in
Fig. \ref{CeRuSi_diag}, where $\delta$ is either the
La-doping content $x$ or the pressure $p$. At $H=0$ a magnetic
singularity separates the antiferromagnetic and paramagnetic
ground states at $\delta_c$. This corresponds to an effective negative pressure $p_c=-3$~kbar or a La-doping $x_c=7.5$~$\%$ applied on the parent compound CeRu$_2$Si$_2$ (Ref. \onlinecite{haen96} and \onlinecite{knafo09}). $\delta_v$ corresponds to a pressure $p_v\simeq2-5$~GPa applied on the parent compound. Above $p_v$ the system is expected to enter into an intermediate valent regime \cite{payer93}. From the antiferromagnetic phase, application of a magnetic field along $\mathbf{c}$ induces a first-order metamagnetic transition at $H_c$, which leads to a critical magnetic field end-point $H_c^*\simeq4$~T at $\delta_c$ (Ref. \onlinecite{fisher91}). Above $\delta_c$, a sharp pseudo-metamagnetic crossover occurs at a magnetic field $H_m$, which reaches 8~T in pure CeRu$_2$Si$_2$ at ambient pressure. Using inelastic neutron scattering, it has been demonstrated that the crossing through $H_m$ is associated with the collapse of the antiferromagnetic correlations together with an enhancement of the low-energy ferromagnetic fluctuations in a quite narrow field-range \cite{sato01,flouquet04}. The transition at $x_c$ is associated with an enhancement of the antiferromagnetic fluctuations \cite{knafo09}. A key observation is the increase by 50~$\%$ of the average effective mass at $H_m$, by comparison to the zero-field value, as measured by the linear $T$-term $\gamma$ of the specific heat \cite{paulsen90}. Above $H_m$, the effective mass decreases strongly with $H$.

\begin{table}[b]
\caption{Specific heat linear coefficient $\gamma$ at ($H=0,p=0$),
($H_{m,c},p=0$), ($H=0,p_c$), low temperature susceptibility $\chi_0^c$ along $\mathbf{c}$, and ratio $\chi_0^c/\chi_0^a$ of the susceptibilities along $\mathbf{c}$ and $\mathbf{a}$, for
CeRh$_2$Si$_2$ and CeRu$_2$Si$_2$.}
\begin{ruledtabular}
\begin{tabular}{lcc}
&CeRh$_2$Si$_2$&CeRu$_2$Si$_2$\\
\hline
$\gamma(H=0,p=0)$ \footnotesize{(mJ/mol.K$^2$)}&23 \footnotesize{[\onlinecite{graf97}]}&350 \footnotesize{[\onlinecite{fisher91}]}\\
$\gamma(H_{m,c},p=0)$ \footnotesize{(mJ/mol.K$^2$)}&40 \footnotesize{[\onlinecite{demuer09}]}&550 \footnotesize{[\onlinecite{paulsen90}]}\\
$\gamma(H=0,p_c)$ \footnotesize{(mJ/mol.K$^2$)}&80 \footnotesize{[\onlinecite{graf97}]}&600 \footnotesize{[\onlinecite{raymond10}]}\\
$\chi_c^0$ \footnotesize{(10$^{-3}$ emu/mol)}&3 \footnotesize{[\onlinecite{mori99}]}&36 \footnotesize{[\onlinecite{haen92}]}\\
$\chi_c^0/\chi_a^0$&1-5 \footnotesize{[\onlinecite{mori99}]}&10-20 \footnotesize{[\onlinecite{haen92}]}\\
\end{tabular}
\end{ruledtabular}
\label{table}
\end{table}

Table \ref{table} recapitulates for CeRh$_2$Si$_2$ and CeRu$_2$Si$_2$, the residual values of the linear $T$-term $\gamma$ of the specific heat at ($H=0,p=0$), ($H_{m,c},p=0$), and ($H=0,p_c$) (for CeRu$_2$Si$_2$, $p_c$ corresponds to a La-doping $x_c=7.5~\%$). The value of the initial susceptibility $\chi_c^0$ along the easy axis $c$ as well as the ratio $\chi_c^0/\chi_a^0$ of the susceptibilities $\chi_c^0$ and $\chi_a^0$ along $c$ and along $a$, respectively, are also summarized in Table \ref{table}. Typically, $\gamma$ and $\chi_c^0$ are 10 times bigger in CeRu$_2$Si$_2$ (Ref. \onlinecite{fisher91} and \onlinecite{haen92}) than in CeRh$_2$Si$_2$ (Ref. \onlinecite{graf97} and \onlinecite{mori99}), which indicates the presence of more intense low-temperature magnetic fluctuations. This leads to an effective mass 10 times bigger in CeRu$_2$Si$_2$ than in CeRh$_2$Si$_2$. The anisotropy of the magnetic susceptibility, estimated via the ratio $\chi_c^0/\chi_a^0$, is also 10 times bigger in CeRu$_2$Si$_2$ than in CeRh$_2$Si$_2$. The proximity of CeRh$_2$Si$_2$ to a valence instability could explain both the reduction of the magnetic anisotropy, via a Kondo broadening/overlapping of the crystal-field levels, and the reduction of the saturated moment of this system \cite{settai97}. In such a scenario, the stronger magnetic fluctuations in CeRu$_2$Si$_2$ could be a consequence of a stronger magnetic anisotropy. Finally, $\gamma(H_{m,c},p=0)$ and $\gamma(H=0,p_c)$ are comparable in both CeRu$_2$Si$_2$ and CeRh$_2$Si$_2$ (see Table \ref{table}). This indicates that the mechanisms which control the  magnetic field- and pressure-enhancements of $\gamma$, and thus of the average effective masses (within a Fermi liquid picture) might be closely connected. This conclusion is compatible with the variations of $A/A_{max}$ as a function of $(p-p_c)/p_c$ and $(H-H_c)/H_c$ reported in Fig. \ref{AHAp}.

 In Fig. \ref{AHcomp}, a comparison is made, for CeRh$_2$Si$_2$ and CeRu$_2$Si$_2$ between the variations of $A/A_{max}$ with the reduced magnetic field $(H-H^*)/H^*$ when $\mathbf{H}\parallel\mathbf{c}$ ($H^*=H_c$ for CeRh$_2$Si$_2$ and $H^*=H_m$ for CeRu$_2$Si$_2$). Below $H^*$, $A/A_{max}$ increases faster with $(H-H^*)/H^*$ in CeRh$_2$Si$_2$ than in CeRu$_2$Si$_2$. Oppositely, above $H^*$ $A/A_{max}$ decreases faster in CeRu$_2$Si$_2$ than in CeRh$_2$Si$_2$. In a conventional magnetic fluctuations scenario \cite{hertz76,millis93,moriya95}, an
enhancement of the magnetic order parameter fluctuations controls
both $A$ and $\gamma$, which leads to a strong effective mass
related to a fixed Kadowaki-Woods ratio $A/\gamma^2$ (Ref. \onlinecite{kadowaki86}). In real systems, $A$ and $\gamma$ can be sensitive to other effects, such as the nature of the magnetic fluctuations (antiferromagnetic, ferromagnetic, single-site etc.) and departures from a unique Kadowaki-Woods ratio can be the
consequence of a wavevector dependence of the magnetic
fluctuations. In high magnetic fields, the persistence of the proportionality between $A$ and $\gamma^2$ has been verified for
CeRu$_2$Si$_2$ (Ref. \onlinecite{daou06}). The differences between CeRu$_2$Si$_2$ and CeRh$_2$Si$_2$ shown in Fig. \ref{AHcomp} might be connected to the differences of their magnetic fluctuation spectra. In both systems, the low-field variations of $A(H)$ are believed to be governed by antiferromagnetic fluctuations. Fig. \ref{AHcomp} is compatible with the expectation that antiferromagnetic fluctuations are less important in CeRh$_2$Si$_2$, which is ordered antiferromagnetically with a relatively high $T_N$ than in CeRu$_2$Si$_2$, which is a paramagnet close to an antiferromagnetic instability. As in CeRu$_2$Si$_2$ (ref. \onlinecite{sato01} and \onlinecite{flouquet04}), critical ferromagnetic fluctuations might play a role in CeRh$_2$Si$_2$ for the enhancement of $A$ and $\gamma$ in a narrow field range around $H^*$. High above $H^*$, for CeRh$_2$Si$_2$ $A(H)$ is still enhanced and its slope remains important up to rather high magnetic fields. This might be related to additional energy scales, such as the Kondo temperature.

\begin{table}[t]
\caption{Lattice parameters $a$ and $c$, unit cell volume $V$, and characteristic energy scales $T_N$, $T_K$, $T_{corr}$, and $\Delta_{CF}$ for CeRh$_2$Si$_2$ and CeRu$_2$Si$_2$.}
\begin{ruledtabular}
\begin{tabular}{lcc}
&CeRh$_2$Si$_2$&CeRu$_2$Si$_2$\\

\hline
$T_N$ \footnotesize{(K)}&36&-\\
$T_K$ \footnotesize{(K)}&35 \footnotesize{[\onlinecite{vildosola05}]}&25 \footnotesize{[\onlinecite{knafo09}]}\\
$T_{corr}$ \footnotesize{(K)}&$>80$&50-80 \footnotesize{[\onlinecite{regnault90},\onlinecite{raymond07},\onlinecite{haen87}]}\\
$\Delta_{CF}$ \footnotesize{(K)}&300 \footnotesize{[\onlinecite{vildosola05}]}&200 \footnotesize{[\onlinecite{vildosola05}]}\\
\hline
$a$ \footnotesize{(\AA)}&4.09 \footnotesize{[\onlinecite{severing89}]}&4.19 \footnotesize{[\onlinecite{severing89}]}\\
$c$ \footnotesize{(\AA)}&10.18 \footnotesize{[\onlinecite{severing89}]}&9.78 \footnotesize{[\onlinecite{severing89}]}\\
$V$ \footnotesize{(\AA$^3$)}&170.6 \footnotesize{[\onlinecite{severing89}]}&171.7 \footnotesize{[\onlinecite{severing89}]}\\
\end{tabular}
\end{ruledtabular}
\label{table2}
\end{table}

A comparison of the magnetic energy scales of CeRh$_2$Si$_2$ and CeRu$_2$Si$_2$ is presented in Table \ref{table2}. In CeRu$_2$Si$_2$, antiferromagnetic correlations were evidenced by inelastic neutron scattering up to a temperature $T_{corr}$ estimated between 50~K and 80~K \cite{regnault90,raymond07}. This can be connected to the initial positive slope of the magnetoresistivity observed for $T\leq50$~K \cite{haen87}. Similarly, the initial positive slope of $\rho_{xx}(H)$ observed up to 80~K in CeRh$_2$Si$_2$ (see Fig. \ref{RH}) is believed to be a manifestation of the strength of the antiferromagnetic correlations. We note that, if the Ce ions behave as independent paramagnetic centers, assuming that the orbital magnetoresistance can be neglected, the slope of the magnetoresistivity would be negative. In CeRh$_2$Si$_2$, the onset of antiferromagnetic correlations is established at a temperature $T_{corr}$ higher than 80~K, i.e., higher than in CeRu$_2$Si$_2$. Table \ref{table2} also shows that, while the unit cell volumes of CeRh$_2$Si$_2$ and CeRu$_2$Si$_2$ are similar, their lattice parameters are quite different. The bigger inter-plane distance between the Ce ions could be a reason for stronger RKKY antiferromagnetic correlations in CeRh$_2$Si$_2$. Indeed, the initial decrease of $T_N$ under pressure is mainly driven by the reduction of the lattice parameter $c$. More precisely, the anomaly at $T_N$ in the thermal expansion along $c$ is negative and 6~times bigger than the one along $a$ (see Ref. \onlinecite{settai97} and Section \ref{thexp}). This indicates that the initial slope of $T_N$ versus uniaxial pressure is 6~times bigger for an uniaxial pressure along $c$ than along $a$. At ambient pressure, the Kondo temperature of CeRh$_2$Si$_2$ is generally estimated at around 35~K \cite{severing89,vildosola05}, which is 50~$\%$ higher than in CeRu$_2$Si$_2$ (see Table \ref{table2}). The origin of antiferromagnetism in CeRh$_2$Si$_2$ is related to the strength of $T_{corr}$ by comparison to $T_K$. A smaller $T_{corr}/T_K$ in CeRu$_2$Si$_2$ might explain why this system is paramagnetic. In addition, CeRh$_2$Si$_2$ is probably close to a mixed valence regime. This is indicated by the smallness of its Sommerfeld coefficient $\gamma$, which reaches only 80~mJ/mol.K$^2$ at the critical pressure $p_c$, and of its small pressure dependence (Ref. \onlinecite{graf97}). The value of $\gamma$ at $p_c$ in CeRh$_2$Si$_2$ is rather similar to that of typical mixed-valence compounds, e.g. $\gamma\simeq40$ mJ/mol/K$^2$ in CeSn$_3$ (Ref. \onlinecite{takke81}). Also, the drop of the magnetic anisotropy in CeRh$_2$Si$_2$ (Ref. \onlinecite{mori99}) might be related to a broadening of the crystal-field levels due to a strong enhancement of $T_K$ when entering into the valence intermediate regime. A unique property of CeRh$_2$Si$_2$ is that its ordering temperature is strongly pressure dependent since the rather high value of $T_{N}=36$~K at ambient pressure is driven to zero at a "relatively small" pressure $p_c=11$~kbar. The strong pressure-dependence of $T_N$ close to $p_c$ could be due to a strong increase of $T_K$ under pressure because of the proximity to a valence transition $p_v$. Knowing that $H_c$ reaches 36~T at $p_c$ in CeRh$_2$Si$_2$ (Ref. \onlinecite{hamamoto00}) one can further speculate if, for pressure bigger than $p_c$ and $p_v$, a magnetic field could push the system to a valence critical point (as discussed in Ref. \onlinecite{watanabe09}). In such case, a study of the variations of $A$ approaching and through the field-induced valence critical point would give important clues about the formation of quasiparticules in heavy-fermion and intermediate-valent systems.

\section{Conclusion}

A study of the heavy-fermion antiferromagnet CeRh$_2$Si$_2$ has
been performed in high magnetic fields of up to 50~T. From resistivity, torque, magnetostriction, and thermal expansion measurements we deduced its magnetic field-temperature phase diagram. It is composed of at least three distinct antiferromagnetic phases and possibly of a tetra-critical point. Fits of our resistivity data showed i) a large $H$-window where the quadratic coefficient $A$ is enhanced, in contrast to the sharpness of the metamagnetic transitions, and ii) that similar values are obtained for $A(p=0)/A(p_c)$ and $A(H=0)/A(H_c)$. This implies that both pressure- and magnetic field-induced criticalities might be controlled by common features, although they are expected to be governed by antiferromagnetic and ferromagnetic fluctuations, respectively. For both pressure- and field-induced magnetic instabilities the effective mass is not found to diverge. Finally, the drop of the resistivity observed at $H_c$ is compatible with the recent observation of a Fermi surface reconstruction at $H_c$, possibly related to a large decoupling between the majority and minority spin band.

\section*{Acknowledgments}

We thank M. Nardone, A. Zitouni, and J. B\'{e}ard for experimental
support, A. Demuer and I. Sheikin for discussions and for
showing us data prior to publication, and L. Malone for carefully reading the manuscript. This work was supported by the French ANR Delice and by Euromagnet II via the EU contract RII3-CT-2004-506239.


\begin{thebibliography}{20}


\bibitem{ohashi03} M. Ohashi, G. Oomi, S. Koiwai, M. Hedo, and
Y. Uwatoko, Phys. Rev. B {\bf68}, 144428 (2003).

\bibitem{settai97} R. Settai, A. Misawa, S. Araki, M. Kosaki, K. Sugiyama, T. Takeuchi, K. Kindo, H. Haga, E. Yamamoto, and Y. \={O}nuki, J. Phys. Soc. Jpn. {\bf66}, 2260 (1997).

\bibitem{abe97} H. Abe, H. Suzuki, H. Kitazawa, T. Matsumo, and G.
Kido, J. Phys. Soc. Jpn. {\bf66}, 2525 (1997).

\bibitem{graf98} T. Graf, M.F. Hundley, R. Modler, R. Movshovich, J.D. Thompson, D. Mandrus, R.A. Fisher, and N.E. Phillips, Phys. Rev. B {\bf57}, 7442 (1998).

\bibitem{kawarazaki00} S. Kawarazaki, M. Sato, Y. Miyako, N. Chigusa, K. Watanabe, N. Metoki, Y. Koike, and M. Nishi, Phys. Rev. B {\bf61}, 4167 (2000).

\bibitem{araki01} S. Araki, R. Settai, T.C. Kobayashi, H. Harima, and Y. \={O}nuki, Phys. Rev. B {\bf64}, 224417 (2001).

\bibitem{movshovich96} R. Movshovich, T. Graf, D. Mandrus, J. D. Thompson, J. L. Smith, and Z. Fisk, Phys. Rev. B {\bf53}, 8241 (1996).

\bibitem{araki02} S. Araki, M. Nakashima, R. Settai, T. Kobayashi, and Y. \={O}nuki, J. of Phys.: Condens. Matter {\bf14}, L377 (2002).

\bibitem{notetransitions} A first-order transition is characterized by a step-like anomaly in the first-derivatives of the free energy, as the entropy, the length, and the magnetization. Subsequently, it is accompanied by a symmetric and sharp extremum, at which the transition temperature or field can be defined, in the second-derivatives of the free energy, as the specific heat, the thermal expansion and magnetostriction coefficients, and the field-derivative of the magnetization (or torque). A second-order transition is characterized by a step-like anomaly in the second-derivatives of the free energy - the transition temperature or field can be defined at the extremum of slope, generally at the half of the step.

\bibitem{araki98} S. Araki, A. Misawa, R. Settai, T. Takeuchi, and Y. \={O}nuki, J. Phys. Soc. Jpn. {\bf67}, 2915 (1998).

\bibitem{villaume07} A. Villaume, D. Aoki, Y. Haga, G. Knebel, R.
Boursier, and J. Flouquet, J. Phys.: Condens. Matter {\bf20},
015203 (2007).

\bibitem{boursier05} R. Boursier, PhD-thesis, Universit\'{e} Joseph
Fourier, Grenoble (2005).

\bibitem{demuer09} F. Levy, A. Demuer, I. Sheikin et al, to be published.

\bibitem{ohkawa90} F.J. Ohkawa, Phys. Rev. Lett. {\bf64}, 2300 (1990).

\bibitem{kadowaki86} K. Kadowaki and S.B. Woods, Solid State Commun. {\bf58}, 507 (1986).

\bibitem{haen96} P. Haen, F. Lapierre, J. Voiron, and J. Flouquet, J. Phys. Soc. Jpn {\bf65} (Suppl. B), 27 (1996).

\bibitem{knafo09} W. Knafo, S. Raymond, P. Lejay, and J. Flouquet,
Nature Physics {\bf5}, 753 (2009).

\bibitem{fisher91} R.A. Fisher, C. Marcenat, N.E. Phillips, P. Hean, F. Lapierre, P. Lejay, J. Flouquet, and J. Voiron, J. Low Temp. Phys. {\bf84}, 49 (1991).

\bibitem{sato01} M. Sato, Y. Koike, S. Katano, N. Metoki, H. Kadowaki, and S. Kawarasaki, J. Phys. Soc. Jpn. {\bf70} Suppl. A, 118 (2001).

\bibitem{flouquet04} J. Flouquet, Y. Haga, P. Haen, D. Braithwaite,
G. Knebel, S. Raymond, and S. Kambe, J. Magn. Magn. Mat.
{\bf272-276}, 27 (2004).

\bibitem{payer93} K. Payer, P. Haen, J.-M. Laurant, J.-M. Mignot, an J. Flouquet, Physica B {\bf186-188}, 503 (1993).

\bibitem{paulsen90} C. Paulsen, A. Lacerda, L. Puech, P. Lejay, J.L. Tholence, J. Flouquet, and A. de Visser, J. Low Temp. Phys. {\bf81}, 317 (1990).

\bibitem{haen92} P. Haen, F. Lapierre, P. lejay, and J. Voiron, J. Magn. Magn. Mat. {\bf116}, 108 (1992).

\bibitem{graf97} T. Graf, J.D. Thompson, M.F. Hundley, R. Movshovich, Z. Fisk, D. Mandrus, R.A. Fisher, and N.E. Phillips, Phys. Rev. Lett. {\bf78}, 3769 (1997).

\bibitem{mori99} H. Mori, N. Takeshita, N. Mori, and Y. Uwatoko, Physica B {\bf259-261}, 58 (1999).

\bibitem{raymond10} S. Raymond, W. Knafo, J. Flouquet, and P. Lejay,
to be published (arXiv:0909.4729v1 [cond-mat.str-el]).

\bibitem{levallois09} J. Levallois, K. Behnia, J. Flouquet, P.
Lejay, and C. Proust, Europhys. Lett {\bf85}, 27003 (2009).

\bibitem{hertz76} J.A. Hertz, Phys. Rev. B {\bf14}, 1165 (1976).

\bibitem{millis93} A.J. Millis, Phys. Rev. B {\bf48}, 7183 (1993).

\bibitem{moriya95} T. Moriya and  T.Takimoto, J. Phys. Soc. Jpn.
{\bf64}, 960 (1995).

\bibitem{daou06} R. Daou, C. Bergemann, and S.R. Julian, Phys. Rev. Lett. {\bf96}, 026401 (2006).

\bibitem{regnault90} L.P. Regnault, J.L. Jacoud, J.M. Mignot, J. Rossat-Mignod, C. Vettier, P. lejay, and J. Voiron, Physica B {\bf163}, 606 (1990).

\bibitem{raymond07} S. Raymond, W. Knafo, J. Flouquet, and P. Lejay, J. Low Temp. Phys., {\bf147}, 215 (2007).

\bibitem{haen87} P. Haen, J. Flouquet, F. Lapierre, P. lejay, J.M. Mignot, A. Ponchet, and J. Voiron, J. Magn. Magn. Mat. {\bf63-64}, 320 (1987).

\bibitem{vildosola05} V. Vildosola, A.M. Llois, and M. Alouani, Phys. Rev. B {\bf71}, 184420 (2005).

\bibitem{severing89} A. Severing, E. Holland-Moritz, and B. Frick, Phys. Rev. B {\bf39}, 4164 (1989).

\bibitem{takke81} R. Takke, M. Niksch, W. Assmus, B. L\"{u}thi, R. Pott, R. Schefzyk, and D.K. Wohlleben, Z. Phys. B {\bf44}, 33 (1981), J.R. Cooper, C. Rizzuto, G. Olcese, J. Phys. (Paris) C {\bf1-32}, 1136, (1971).

\bibitem{hamamoto00} H. Hamamoto, K. Kindo, T.C. Kobayashi, Y.
Uwatoko, S. Araki, R. Settai, and Y. Onuki, Physica B
{\bf281-282}, 64 (2000).

\bibitem{watanabe09} S. Watanabe, A. Tsuruta, K. Miyake, and J.
Flouquet, J. Phys. Soc. Jpn. {\bf78}, 104706 (2009).


\end{thebibliography}
\end{document}